\documentclass[12pt]{article}
\usepackage{graphicx}% Include figure files

\usepackage{epsfig}
\usepackage{latexsym}
\usepackage{amsmath}

\begin{document}

\title{   %Relations between coauthor core measures for publications in peer review  journals and in proceedings
% Surprising or unexpected 
Effect of religious rules on time of conception in Romania from 1905 to 2001
% \\  \\Coauthors  are a "positive measure"  of a principal investigator role  for publications in peer review  journals and in proceedings %Measure of the
%$m_a^{(j)} +   0.41\;m_a^{(p)} =  m_a^{(jp)} $   \\   and\\
%$A_a^{(j)} + 1.36\; A_a^{(p)}  = A_a^{(jp)} $
%\\ mav03
}

\author{Claudiu Herteliu$^1$, Bogdan Vasile Ileanu$^1$, Marcel Ausloos$^{2,3,4}$, Giulia Rotundo$^5$ %$^{a,b}$
%\footnote{  previously at  GRAPES@SUPRATECS, ULG, B5a Sart-Tilman,  B-4000 Li\`ege, Euroland;  e-mail address: marcel.ausloos@ulg.ac.be}
 \\ %$^a$
%$^a$R\'es. Beauvallon, r. Belle Jardini\`ere, 483, \\ B-4031 Liege, Wallonia-Brussels Federation\\ and\\
% $^b$
 %eHumanities group
\\Ê$^{1}$The Bucharest University of Economic Studies,  Bucharest, Romania    \\Ê\\Ê$^{2}$School of Management, University of Leicester, \\ University Road, Leicester  LE1 7RH, UK  \\ \\ $^{3}$GRAPES, r. Belle Jardini\`ere, 483, \\ B-4031 Li\`ege, Wallonia-Brussels Federation \\ \\Ê$^{4}$Royal Netherlands Academy of Arts and Sciences\\
Joan Muyskenweg 25, 1096 CJ Amsterdam, The Netherlands 
\\ \\
$^{5}$ Sapienza University of Rome, Faculty of Economics,
\\Department of Methods and models for Economics, Territory and Finance,
%Dipartimento di Metodi e modelli per l'economia, il territorio e la finanza
\\via del Castro Laurenziano 9, I-00161 Roma, Italia 
 }

 \date{\today}
\maketitle
 \vskip 0.5 cm

\begin{abstract}
 Population growth (or decay) in a country can be due to various f socio-economic constraints, as demonstrated in this paper. For example, sexual intercourse  is banned in various religions, during Nativity and Lent fasting periods. Data consisting of registered daily birth records for very long (35,429 points) time series and many (24,947,061) babies in Romania between 1905 and 2001 (97 years) is analyzed. The data was obtained from the 1992 and 2002 censuses, thus on persons alive at that time.  
 We grouped the population into two categories (Eastern Orthodox and Non-Orthodox) in order to distinguish religious constraints and performed extensive data analysis in a comparative manner for both groups. From such  a long time series data analysis, it seems that the Lent fast has a more drastic effect than the Nativity fast over baby conception within the Eastern Orthodox population, thereby differently increasing the population ratio. Thereafter, we  developed  and tested econometric models where the dependent variable is the baby conception deduced day, while the independent variables are: (i) religious affiliation; (ii) Nativity and Lent fast time intervals; (iii) rurality; (iv) day length; (v) weekend, and (vi) a trend background.   Our findings are concordant with other papers, proving differences between religious groups on conception, - although reaching different conclusions  regarding the influence of weather on fertility.  The approach seems a useful hint  for developing econometric-like models in other sociophysics prone cases.
  \end{abstract}

  \section{Introduction  }\label{sec:intro}

Babies are born and therefore conceived non-uniformly over the year. The world-wide trend is
affected by seasonality (Quetelet, 1826; Lam and Miron, 1991; Cancho-Candela et al., 2007). The determinant factors producing these seasonal effects can be grouped (Friger et al., 2009)
into so called ÒnaturalÓ characteristics (latitude, weather conditions, day-length (DL)) and ÒartificialÓ
ones (demographic, economic, socio-cultural, including religion).

The seasonal effects of natural factors have been extensively studied. The target populations
are spread all around the world (Lam and Miron, 1994; Martinez-Bakker et al., 2014; Friger et
al., 2009; Dor\' elien, 2013; Hubert, 2014), but very rarely were they focused on Romania (except
for Huber and Fieder, 2011). The findings on climate effects vary across time and geographical
location. They have been extensively analyzed, the importance attributed to them ranging from
high (Villerme, 1831) to Òas unimportantÓ in e.g. some French population (Regnier-Loilier and
Rohrbasser, 2011) or seen as acting jointly with economic development (Huber and Fieder,
2011; Seiver, 1985) due to the reasoning that poor populations are negatively affected by low
temperatures during winter or, on the contrary, by hot summers without air conditioning.
Other researchers (Macfarlane, 1974; Lam and Miron, 1994) stated that weather affects the
seasonality of births as long as there are major temperature differences over the year. Other
natural factors like day-length (Bronson, 2004; Rojansky et al., 1992) or moon phases (Criss and
Marcum, 1981; Jongbloet, 1983) were also tested for their possible connection to birth
seasonality.

From the artificial factors which were recorded as having some influence over the seasonality
of births, let us mention the demographic ones: age, marital status, education,
intergenerational effects (Bobak and Gjonca, 2001; Huber and Fieder, 2011) and economic
ones: income, rurality (Kestenbaum, 1987; Vitzthum et al., 2009), as well as socio-cultural
characteristics: religion, ethnicity (Friger et al., 2009; Hubert, 2014). It turns out occurs that
research on the influence of the artificial factors is rare, perhaps due to some lack of precise
data. In this context, our paper aims to answer the following question concerning the influence
of religiosity on the seasonality of dates of conception and of birth. More precisely,  we raise questions as: did the
interdiction of sexual intercourse during fast  (Lent and Nativity) periods have any effect on the date of conception in
twentieth century Romania? Has that socio-cultual constraint an effect on population growth?

   \section{Data sample }   \label{sec:dataanal}
 Romania is a country with a large majority (more than 86\% nowadays) of Eastern Orthodox
population as can be observed in  Fig. 1, which is based on censuses performed across
Romanian territories in the last 150 years. Other religious denominations have been present in
different proportions during the last 150 years in regions that are now part of Romania. For
example, it is apparent that Transylvania is the most heterogeneous Romanian territory with
regard to religious diversity (Herteliu, 2013). However, this effect is apt to be minor, and our
current analysis is conducted at the national level, without distinguishing regions.
 
             \begin{figure}
 \centering 
 \includegraphics [height=8.0cm,width=10.0cm]{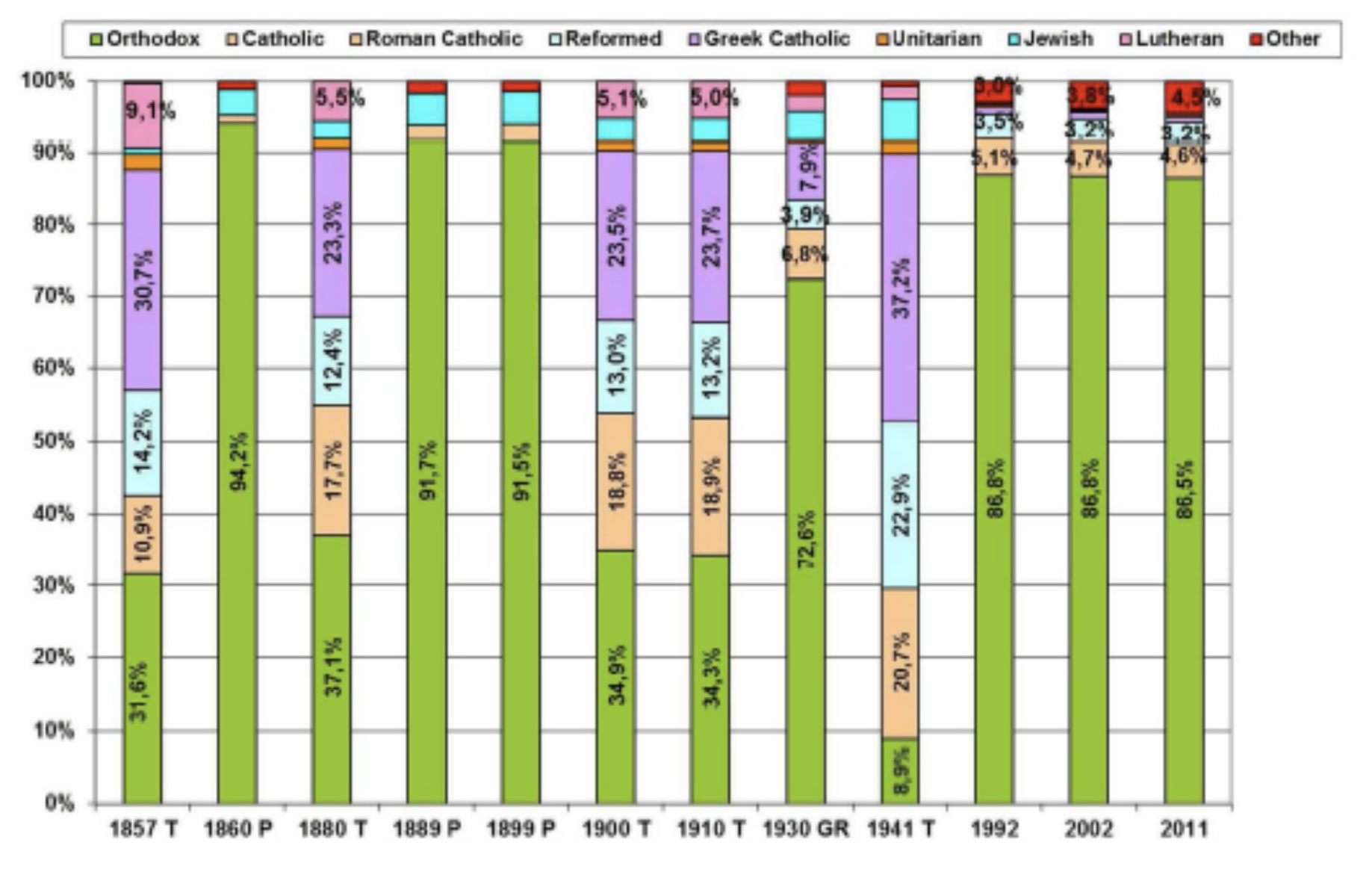}
%Plot10UEFAlili2f.pdf
%  \includegraphics [height=20.0cm,width=16.0cm]{Fig1Plot2colallteams2fits.pdf} \end{figure}
 \caption   { Religious affiliation in Romania
Source of data: Censuses performed across Romanian territories in the last 150 years. Please note that in
communist era four censuses were performed (1948, 1956, 1966 and 1977). Due to political bias, none of
them included the questionnaire items regarding religious affiliation.
Note: The data on Romanian territories are organized or grouped differently
from today. The codification is the following: T Ð Transylvania; P Ð Romanian Principalities
(Moldavia and Wallachia); GR Ð Greater Romania }    \label{figure1}
 \end{figure}

 It is known that the operation of complex mechanisms acting on or related to religion is not an
easy task to investigate (Herteliu, 2007; Herteliu and Isaic-Maniu, 2009; Ileanu et al., 2012). In
this context, to get some statistical balance given the vast Eastern Orthodox majority, the non-
Orthodox segments are grouped together, thus transforming the religious affiliation into a
simple yes or no variable. Even if some other religions grouped into the category Ònon-
OrthodoxÓ have periods of fasting during the year, due to their particularities - such as the use
of a different calendar, different dates for Easter, different rules for fast periods, or other
features specific to various religious sects or denominations, their exclusion from the well
defined Eastern Orthodox group was preferred for the sake of simplicity.
 
It is rather well-known that in the Eastern Orthodox tradition, in addition to abstaining from
particular types of food and drinks during fasting periods, individuals should also avoid sexual
intercourse. Support for these extensive fasting interdictions specific to the Eastern Orthodox
tradition ($http://orthodoxinfo.com/praxis/fasting_sex.aspx$ retrieved in March 2015) are based
on interpretations of certain quotes from the Bible (such as Romans 8:12-14 or Galatians 5:16-
17). One of our goals is to find whether these constraints have been implemented.

 \subsection{ Data sources}

The analysis is based on a very long (35,429 points =  365 (days/year) ? 97
years + 24 leap years) daily time series representing all births (24,
947,061) of persons alive at the 1992 or 2002 censuses (similar to Kestenbaum, 1987). Other
long data series have been studied before (Cancho-Candela et al., 2007; Ausloos et al., 2015; 
Rotundo et al., 2015) using different methodologies. In the present case, data points were
recorded for 01/01/1905 to 31/12/1991 (31,776 points) from the 1992 census, while the other
3,653 points (corresponding to the period 01/01/1992 to 31/12/2001) are from the 2002
census. The data source is the Romanian National Institute of Statistics (NIS) via a query tool
available within NIS's intranet ($http://happy:81/PHC$) regarding 2002 and 1992 censuses. For
convenience, in the current research, the terms Òborn per day" will be used instead of "number
of births on a given day for persons still alive at the 1992 and respectively at 2002 Census". The
data for DL and moon phase (proportion of moon disc illumination) was obtained from
http://www.timeanddate.com. For daily temperatures the source for data is Romanian National
Meteorological Agency (via $http://www.ecad.eu/dailydata/customquery.php$ (retrieved on
December 2013)). The dates for Easter in the Eastern Orthodox religion during the last 110
years is obtained from $http://www.smart.net$ (retrieved on December 2013).

 \subsection{ Variables}
(a) \underline{Variables from censuses }: Data was based on persons' self-declared
birthday. In addition, we use data about: (i) religion: distinguishing Eastern Orthodox (EOx)
and Non-Orthodox (NOx) affiliation. We must assume that the religious affiliation of the
parents is well defined, based on the information given by their offspring at census time, and is
the same for both parents. Even so, the available statistical information showed that the
intensity of changing religious affiliation across the Romanian population is very low (Ileanu et
al., 2012); (ii) rurality: the percentage of people located in rural or non rural areas. This
introduces a double filter: Eastern Orthodox (rurality EOx: rEOx) and Non-Orthodox
(rurality NOx: rNOx), the alternative being Eastern Orthodox  in Urban communities (EOxU) and Non-Orthodox  in Urban communities (NOxU).
 
After subtracting 280 days from the registered birthday, an estimate of the conception day was
obtained (Seiver, 1985; Lam et al., 1996). However, date-of-birth evidence can only point to to
sexual intercourse leading to successful pregnancies. It is known that Lent is 48 days long while
the Nativity fast lasts 40 days. The uncertain conception date, since not every child is born
exactly 280 days after conception (Pana et al., 2015), overlaps the 40 or 48 day fast. We
assume that the birth rate is correlated with the rate of intercourse. Practically, family planning
and contraception methods have not been very common in Romania (it was certainly not
encouraged by the communist regime between 1966 and 1989) (Chelbegean, 2010). We
 assume that the birth rate is correlated with the rate of intercourse. Recall that this {\it a posteriori}
estimate of sexual intercourse resulting in a birth a is made primarily to correlate to ÒnaturalÓ
variables, i.e., weather conditions and day light (or night) duration at the time of the sexual
activity.
 
For each year, the fraction of births $y_i$ on given day, $i$, were computed separately for the
 orthodox $y_i^{EOx}$ , and non-orthodox populations  $y_i^{NOx}$, using:
 \begin{equation}
 y_i^{EOx} =\frac{n_i^{EOx}}{\sum n_i^{EOx}}
 \end{equation}
 and
  \begin{equation}
 y_i^{NOx} =\frac{n_i^{NOx}}{\sum n_i^{NOx}}
 \end{equation}
 where  $n_i^{EOx}$ and $n_i^{NOx}$ is the number of births on the specific day $i$ for EOx and NOx respectively, while $\sum n_i^{EOx}$ and $\sum n_i^{NOx}$
 is the total number of births during that year for the respective populations.
 
 The dependent variables in the regression models are calculated by dividing the daily fraction of
births (1) and (2) by the daily expected (uniform) distribution (UD) across the year which is
$y^{UD} = 1/365$ for non-leap years and $y^{UD} = 1/366$
 for leap years. Therefore, we define the following
two dependent variables:
  \begin{equation}
d_i^{EOx} =\frac{y_i^{EOx}}{y^{UD}}
 \end{equation}
 and
  \begin{equation}
 d_i^{NOx} =\frac{y_i^{NOx}}{y^{UD}},
 \end{equation}
  respectively and using the  appropriate $y^{UD}$.
  
  (b) \underline{Nativity and Lent fast variables}: We used a countdown variable (in days) to Christmas or
Easter for each day of Nativity or Lent fast, instead of a simple binary variable indicating
whether or not conception occurred in the fasting period. We preferred this because it is possible that religious constraints were observed more scrupulously the closer one was to
either Easter or Christmas. This countdown variable was computed having the conception date
as the reference point. Subsequently, Days Before Christmas (DBC) was computed from:
    \begin{equation}
DBC = 0 \; or \;  t
 \end{equation}
 if the conception occurs within or outside Nativity fast time interval; $t $ represents the number of days to Christmas.
 
 Similarly, Days Before Easter (DBE) was defined but where $t$ represents the number of days to Easter:
     \begin{equation}
DBE = 0 \; or \;  t
 \end{equation}
 For completeness, we note that Romania adopted the Gregorian calendar in 1919. Therefore,
the 31st of March 1919 was followed by the 14th of April 1919. Moreover, the Romanian Eastern
Orthodox Church is using an updated form of the Julian calendar. Thus, Christmas is celebrated
in Romania on the 25th of December. However, Easter for the Romanian Eastern Orthodox
Church is usually not synchronized to that of the Roman Catholic Church. Even so, it is
emphasized that these specific calendar change details do not interfere with the results of our
research, since the data regarding registered birthdays in the 1992 and 2002 censuses were
based on the Gregorian calendar.

   (c) \underline{Time interval related variables}: 
   Some binary variables were inserted in order to mark some
particular time windows during which the conception date might have occurred and influenced
by global events or days of the week. These are: (i) First World War (1stWW) taking the value 1
for conception days between 28 July 1914 and 11 November 1918 and 0 outside this timespan;
(ii) Second World War (2ndWW) having a 1 value for conception days between 1 September
1939 and 2 September 1945 and 0 outside this timespan; (iii) weekend's days having a value = 1
if the conception day is on a Saturday or Sunday and 0 for Monday, Tuesday, Wednesday,
Thursday and Friday and (iv) squared trend ($t^2$ for $t$=1 to 35,429).
   
      (d) \underline{Natural variables}:
      The effect of some natural variables is also analyzed through: (i) day length
(DL) measured as the number of minutes of daylight during each 24 hours of a day. This
information refers to Bucharest as the capital city of Romania, but is taken as to be the same
for wherever place in Romania; (ii) temperature in Celsius degrees (mainly measured for
Bucharest)  {\it (Due to the usual
intrinsic error bars on meteorological measurements (and variation across the country and
throughout the day), using Bucharest temperatures is just a rough approximation to local
temperature.)},  and (iii) proportion of the moon disc being illuminated (having also Bucharest as the reference point), within the same approximation.

    \subsection{Statistical analyses}
    
    The analysis began in qualitatively to determine whether births (or conception) vary
throughout the year. Graphs were drawn with various softwares (Kaleidagraph, EViews, Excel,
Power Point). Subsequently, the daily average conception within and outside fasting periods
was computed. Prior to the econometric modeling, we checked a possible periodicity via a Fast
Fourier Transformation performed with SPSS for  $2^{15}$ cases (the last 32,768 days).
    
    The final step in the data analysis consists in the design of several econometric models.
Breaking points in the trend of the dependent variables were identified with Cusum SQ Test
performed with EViews. Regressions were performed with IBM-SPSS (Ordinary Least Square -
OLS estimations), - the same model being run for each phase/cycle (identified after a Cusum SQ
Test). At this stage, factors suspected of being collinear were removed (e.g. day length was
chosen instead of temperature) from the regression models. The same procedure was
performed for factors with low statistical significance (e.g. moon phases).
    
  \section{  Results  }\label{sec:Results}

  \subsection{Preliminary data inspection}
   
        \begin{table} \begin{center}  
\begin{tabular}[t]{|cc|c|c|c|}
\hline  
 \multicolumn{2}{|c|}{Indicator}&EOx&NOx \\ \hline	 				
N	&	Valid	&	366	&	366 	\\					
	&	Missing	&	0	&	0	\\					
Mean	&		&	59219.15	&	9082.180	\\					
Median	&		&	59418.00	&	9050.0000	\\					
Std.Deviation	&		&	8906.85	&	1049.23	\\					
Skewness	&		&	0.93	&	2.21	\\					
Kurtosis	&		&	13.52	&	37.59	\\					
Minimum	&		&	11291	&	1904	\\					
Maximum	&		&	131311	&	20016	\\					
Sum	&		&	21629796 	&	3317625	\\					
Percentiles	&	20	&	53554.80	&	8452.8	\\					
	&	40	&	57996.20	&	8851.8	\\					
	&	60	&	60449.60	&	9209.2	\\					
	&	80	&	65141.40	&	9.618.6	 Ê\\ \hline
	\end{tabular}  
   \caption{Descriptive statistics for daily   aggregated data of  population growth  }\label{TableI}
\end{center} \end{table} 
  As shown by Huber and Fieder (2011) for women born between 1920 and 1955 and older than
45 years (completed fertility), Romanian births register a non-uniform distribution across the
year with a maximum in June and minima in December and January, as obtained from monthly
data.
  
Our raw data contains the number of births on each day of the year (EOx and NOx) for a
long period (97 years). In order to present this extensive data in a concise way, a pivot table
was designed separately containing monthly and daily aggregated values for EOx and
NOx. For both populations, the mean and median are very close to each other (less than
0.4\%) while the standard deviations for the means are quite low (15\% of the mean for EOx
and 11.6\% for NOx). The statistics of both populations are slightly asymmetric and
leptokurtic. Other descriptive statistical characteristics are available in Table \ref{TableI}. Next, the
aggregated date were sorted into quintiles (Dedu et al., 2014; Giuclea and Popescu, 2009)
obtained by quintiles. The following intervals were obtained for EOx (Fig. \ref{TableII5}):  %(figure 2), 
 (q1) days having under 53,554.8 births labeled in red; (q2) days having between 53,554.8 and 57,996.2
births labeled in orange; (q3) days having between 57,996.2 and 60,449.6 births labeled in
yellow; (q4) days having between 60,449.6 and 65,141.4 births labeled in light green and (q5)
days having over 65,141.4 births labeled in dark green. Similarly for NOx  (Fig. \ref{TableII6}), %(figure 3), 
 the quintile intervals are (q1) days having under 8,452.8 births (red); (q2) days having between 8,452.8 and 8,851.8 births (orange); (q3) days having between 8,851.8 and 9,209.2 births
(yellow); (q4) days having between 9,209.2 and 9,618.6 births (light green) and (q5) days having
over 9 618.6 births (dark green).

  This sorting by quintiles of almost one century's data can (by avoiding compensation through
averaging over unequal size month) provide some statistically significant differences among
daily distributions across the year. It can easily be seen that this compensation does not
presently occur is our case. Therefore, it may be deduced that the outlined monthly or daily
differentiation does not lead to a random outcome.

       \begin{table}
 \begin{center}  
 \includegraphics [height=5.0cm,width=10.0cm]{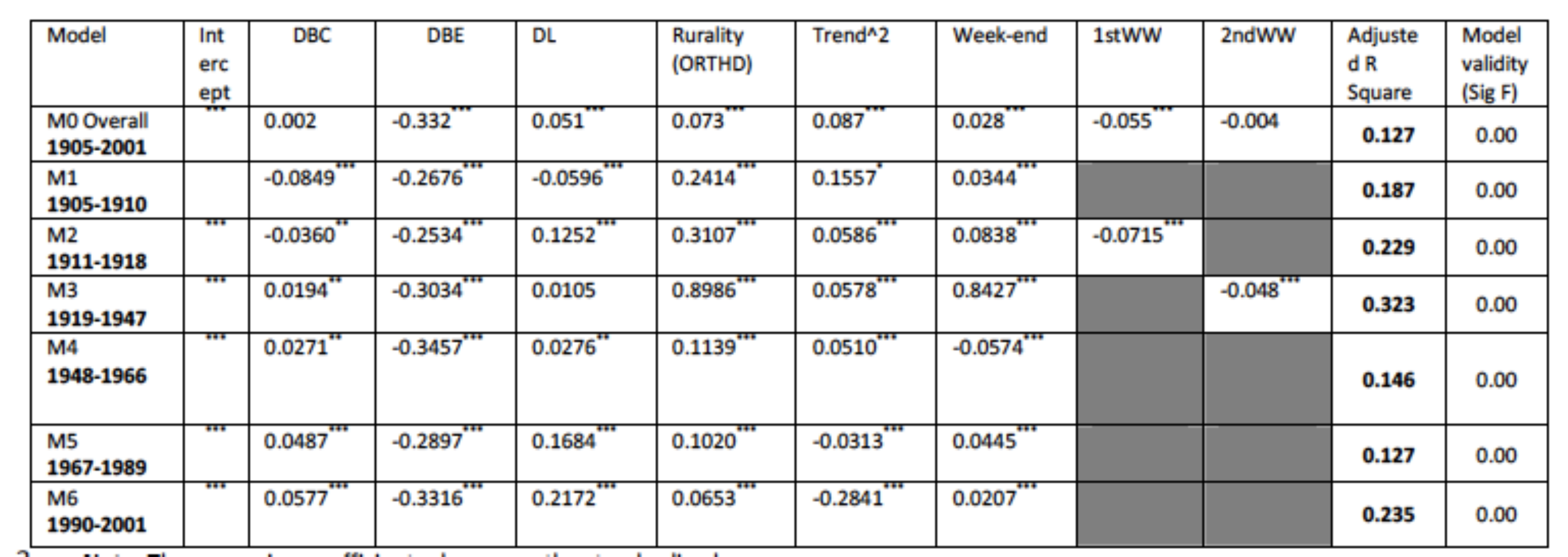}
%Plot10UEFAlili2f.pdf
%  \includegraphics [height=20.0cm,width=16.0cm]{Fig1Plot2colallteams2fits.pdf} \end{figure}
 \end{center}  \caption   { %Plot 6m2m35
The logarithm of the ratio of conceived orthodox babies relative to assumed uniform daily value for 1 year.
 The regression coefficients shown are the standardized ones (prior to estimation all independent variables are standardized). Factors (independent variables) are: Days Before Christmas (DBC), Days Before Easter (DBE), Day Length in
minutes (DL), Rurality for ORTHD, squared trend ($t^2$ for $t$ = 1 to 35,429), Binary variables: (i) FirstWorldWar (1stWW)taking the value 1 for conception days between 28 July 1914 and 11 November 1918 and 0 outside this timespan; (ii) Second
WorldWar (2ndWW) having a 1 value for conception days between 1 September 1939 and 2 September 1945 and 0 outside this timespan; (iii) weekend days having a value   1 if the conception day is on a Saturday or Sunday and 0 for Monday,
Tuesday,Wednesday, Thursday and Friday. Bold values denote the modelÕs goodness of fit, adjusted by number of independent variables.
 $^{*}$Statistically significant at level 10\%; $^{**}$Statistically significant at level 5\%;  
 $^{***}$Statistically significant at level 1\%.   }  
  \label{TableII5}
 \end{table}  
 
      \begin{table}
 \centering 
 \includegraphics [height=5.0cm,width=10.0cm]{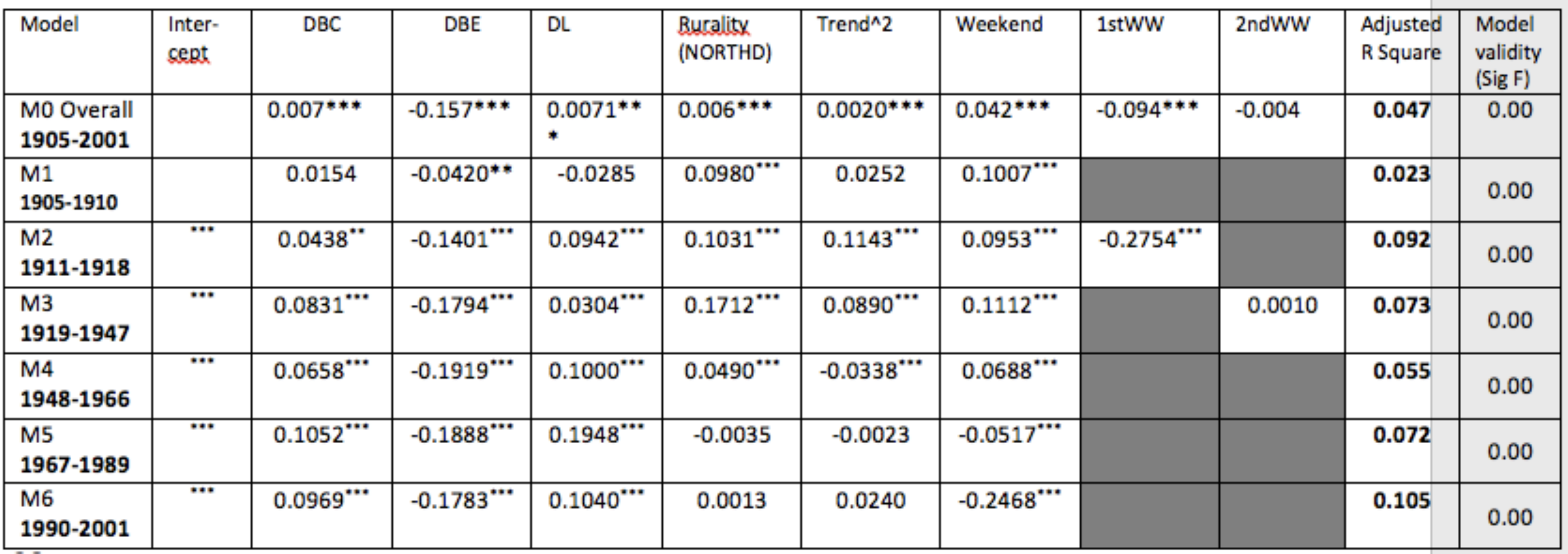}
%Plot10UEFAlili2f.pdf
%  \includegraphics [height=20.0cm,width=16.0cm]{Fig1Plot2colallteams2fits.pdf} \end{figure}
 \caption   { %Plot 6m2m35
The logarithm of the ratio of conceived Non-Orthodox babies relative to assumed uniform daily value for 1 year.
The regression coefficients shown are the standardized ones (prior to estimation all independent variables are standardized). Factors (independent variables) are: Days Before Christmas (DBC), Days Before Easter (DBE), Day Length in
minutes (DL), Rurality forNORTHD, squared trend ($t^2$ for $t=$ 1 to 35,429), Binary variables: (i) FirstWorldWar (1stWW)taking the value 1 for conception days between 28 July 1914 and 11 November 1918 and 0 outside this timespan; (ii) Second
WorldWar (2ndWW) having a 1 value for conception days between 1 September 1939 and 2 September 1945 and 0 outside this timespan; (iii) weekend days having a value = 1 if the conception day is on a Saturday or Sunday and 0 for Monday,
Tuesday,Wednesday, Thursday and Friday. Bold values denote the modelÕs goodness of fit, adjusted by number of independent variables.
**Statistically significant at level 5\%; 
*** Statistically significant at level 1\%. }    \label{TableII6}
 \end{table}

         \begin{figure}
 \centering 
 \includegraphics [height=10.0cm,width=10.0cm]{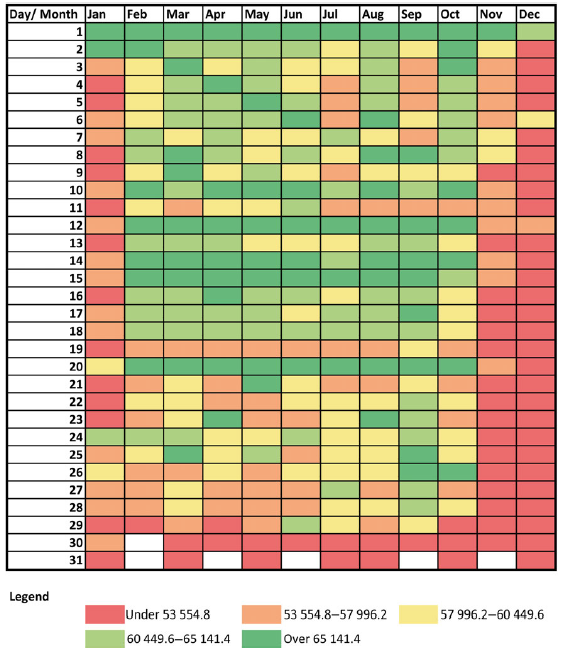}%figure2.pdf}
%Plot10UEFAlili2f.pdf
%  \includegraphics [height=20.0cm,width=16.0cm]{Fig1Plot2colallteams2fits.pdf} \end{figure}
 \caption   { %Plot 6m2m35
%Dependent variable: ln(EOx): the logarithm of the ratio of   conceived   orthodox babies relative to and assumed uniform daily value for one year 
Quintile distribution of daily aggregated births for the Eastern Orthodox (EOx) population during 1905-2001.}    \label{TableII}
 \end{figure}  
 
         \begin{figure}
 \centering 
 \includegraphics [height=10.0cm,width=10.0cm]{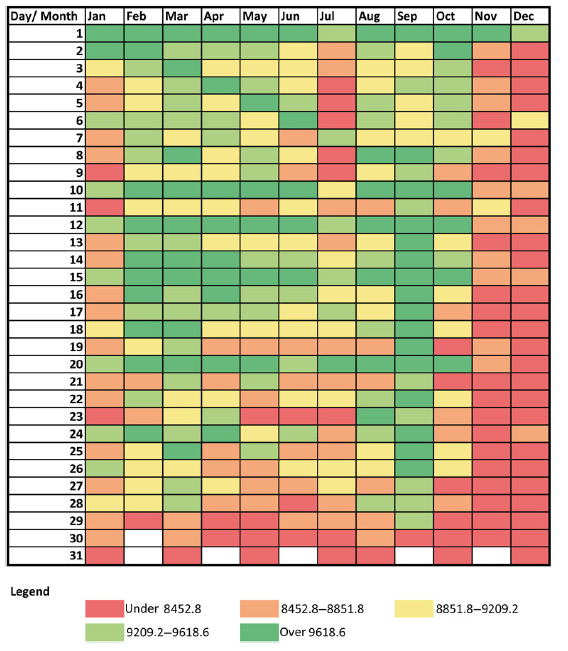}
 %figure3.pdf}
%Plot10UEFAlili2f.pdf
%  \includegraphics [height=20.0cm,width=16.0cm]{Fig1Plot2colallteams2fits.pdf} \end{figure}
 \caption   { %Plot 6m2m35
%Dependent variable: ln(NOx): the logarithm of the ratio of   conceived non-orthodox babies relative to and assumed uniform daily value for one year
Quintile distribution of daily aggregated births for the Non-Orthodox (NOx) population during 1905-2001. }    \label{TableIII}
 \end{figure}  
 
  Referring to Fig 2, for the EOx population, there are important differences between
November, December and January and the rest of the year, and also between the first two
thirds of a month compared to the last third. These monthly differences were tested by a  $\chi^2$ 
  test and are statistically significant (computed $\chi^2$  was 131,285.4 being
significant with a p value lower than 0.001). Again referring to Fig. 2, the lowest aggregate
number of births for a specific day was, naturally, on 29th of February (11,291 births) while the
maximum was on the 1st of January (131,311 births). We will comment on the large number of
births on 1 Jan below.
   
   In the case of the NOx population, (See Fig. 3) the lower birth figures in November and
December are consistent with lower number of births in those months in the EOx
population, while January is not so distinct as in the NOx case. The distribution of the daily
figures outside the November-January time span seems to be more equally balanced for
NOx, - no persistent pattern is visible, in contrast to the EOx case. Similarly in the
EOx and the NOx population, the minimum aggregate number of births was registered
on 29th of February (1,904 births) while the higher one occurs on 1st of January (20 016 births).
For this group, the computed $\chi^2$  was 9,184.2 showing that the differences are
statistically significant with a p value lower than 0.001.
      
  For both (EOx and NOx) populations, Figs 2- 3 show a noticeably high number of
births (almost dark green) on the 1st day of any month and a low number of births) on the last
day of each month.
  
As noted above for both populations, there is an anomalously high number of births on Jan. 1st.
In fact there are twice as many births on Jan 1st than on Jan 2nd and almost 4 times as many births
as on the 31st of December. We call this anomaly the Ò1st of January effectÓ; further work
should be done on this subject. A similar effect was noted by MacFarlane (1974). In brief,
possible explanations for this effect might be: (i) parents' psychological comforting thought that
their child is considered to be one year younger if registered on January 1st instead of
December 31st (a quite comfortable situation for both girls and boys since there would be a
delay of one year to the compulsory military training, - effective in Romania prior to 2007); (ii)
municipal recorders not working during winter holidays therefore increasing the likelihood of
an incorrect record (despite the fact that children can be registered on the 5th of January
mentioning the correct birth date); (iii) on the 1st of January there is a very important holiday
for the EOx (Saint Greater Basil), evidenced by many children born around this day named
Basil. (According to $http://www.capital.ro/peste-600000-romani-isi-serbeaza-ziua-onomasticade-
sfantul-vasile-142714.html$, retrieved on December 2014, there are almost 600 000 people
in Romania named Basil or the female version); and (iv) mistaken registration. (Even if it is not statistically important, one co-author of the current paper has a brother registered as born on
January 1st, even if the event certainly occurred, on 2nd January 1:00 AM).
  
  Since the variation of the number of births across the year is undoubtable, we now address the
primary concern of this research: did major fast periods (i.e. before Christmas and Easter) affect
conception or not ?
First, note that several religions (Roman Catholic, Greek Catholic, Old Rite Eastern Orthodox)
grouped within NOx category still have different Lent and Nativity fast periods. Moreover,
the fast periods (especially for Lent) are usually not similar to EOx. Therefore, a visual
inspection of this variability (Figs. 4 - 5) was conducted, only for EOx. Prior to this, it
seems useful to compute for each year the daily average conceptions within such fast (or out of
fast) periods.
  
         \begin{figure}
 \centering 
 \includegraphics [height=8.0cm,width=10.0cm]{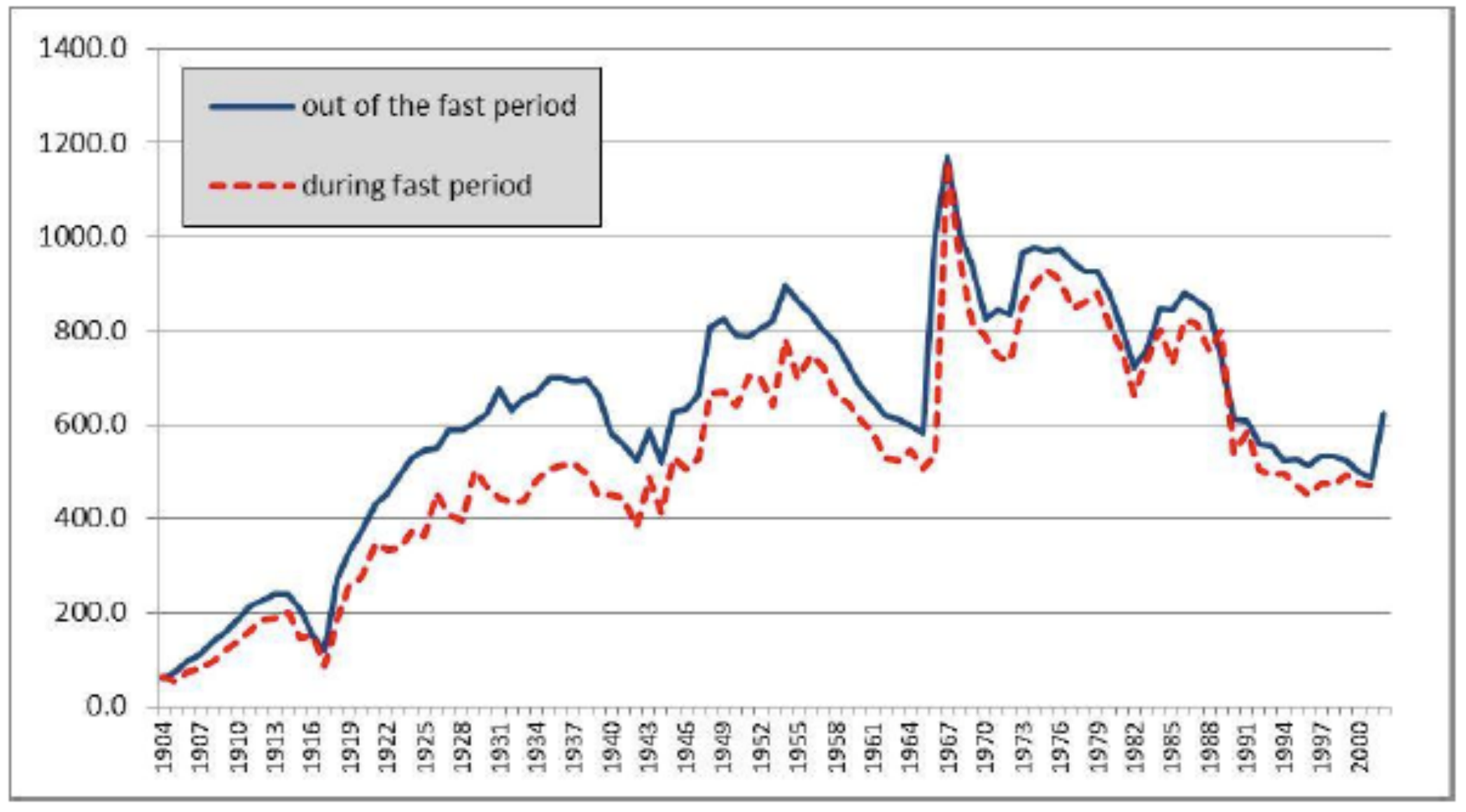}
%Plot10UEFAlili2f.pdf
%  \includegraphics [height=20.0cm,width=16.0cm]{Fig1Plot2colallteams2fits.pdf} \end{figure}
 \caption   { %Plot 6m2m35
Daily average   number of conceptions during Lent fast/out of fast period for Eastern
Orthodox (EOx) }    \label{figure4}
 \end{figure}  
        \begin{figure}
 \centering 
 \includegraphics [height=8.0cm,width=10.0cm]{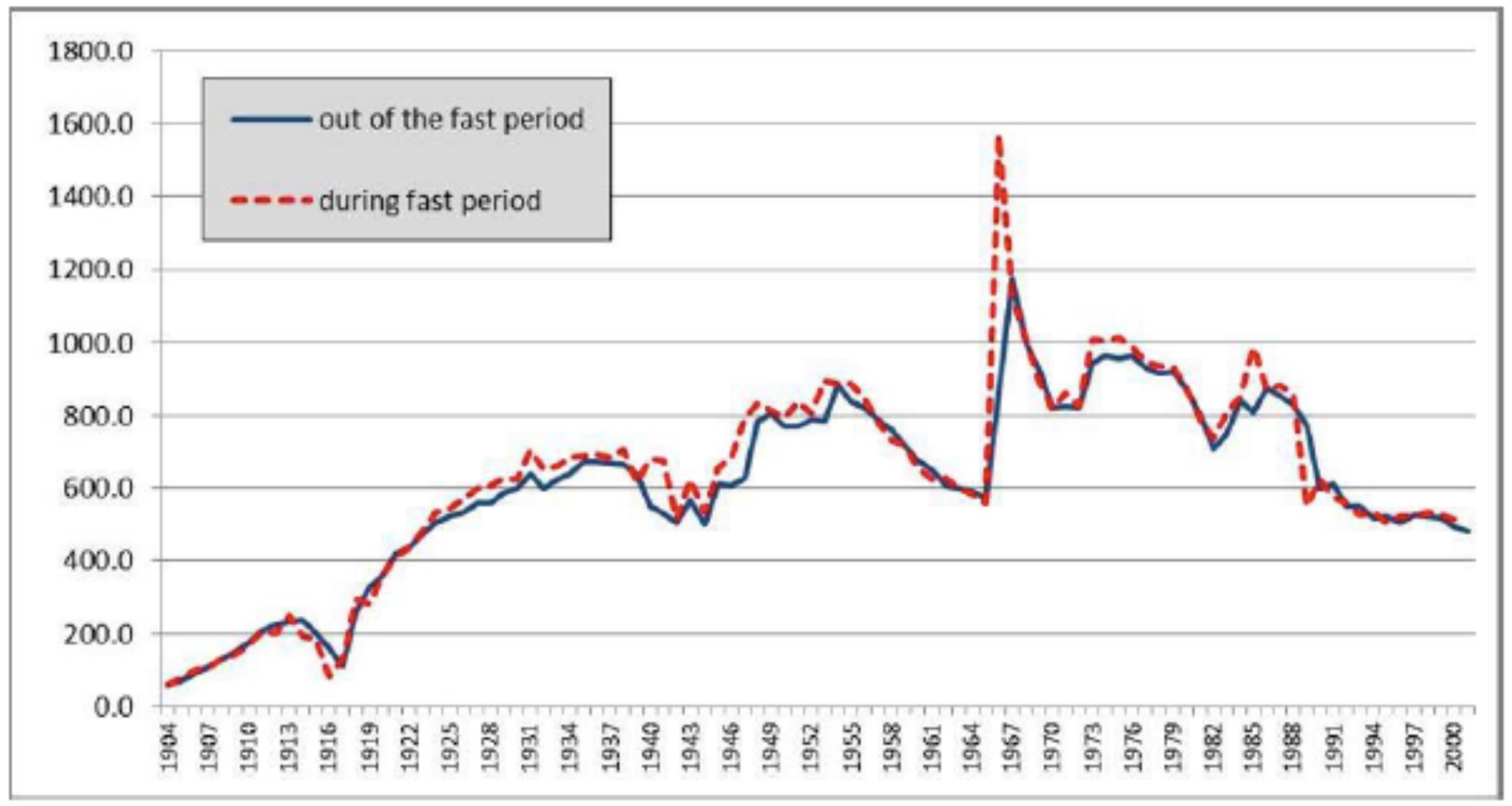}
%Plot10UEFAlili2f.pdf
%  \includegraphics [height=20.0cm,width=16.0cm]{Fig1Plot2colallteams2fits.pdf} \end{figure}
 \caption   { %Plot 6m2m35
Daily average number of conceptions during Nativity fast/out of fast period for Eastern
Orthodox (EOx) }    \label{figure5}
 \end{figure}  
 
Despite the long time series which certainly must depend on many local or general factors,
Fig. 4 shows that the daily mean number of conceived children by orthodox is significantly
lower during Lent than during the rest of the year. In the whole range of almost 100 years only
two exceptions to this rule can be seen: (i) during the First World War (more exactly, in 1916)
and (ii) during 1989, a year which had a strong emotional impact on the Romanian population,
when there was a revolution and a major change of political system. Even if births increased
because of Ceausescu's decree, the number of conceptions Òout of fastÓ is still larger than the
number of conceptions during fasting periods. Thus, even if in the short term (1967-1968) the
Ceausescu Decree reduced the gap between "out of fast" and "during the fast" on the long term, the decree had not enough force to completely eliminate the effect of ecclesiastical
admonitions.

Similar plots of conception within and without the Nativity fast (Fig. 5) show dramatically
different behavior. In this case the rule of Òfewer conceptions during fast periodsÓ was broken
in 67 from all 97 available years.
To complete the basic statistical analysis, an enhanced econometric analysis is described and
performed in the Econometric models sub-section,  below.
  
   \subsection{Identification of periodicity}

The analytical approach of our paper is an econometrical one. However, before applying
econometric model, it is useful to test whether the main dependent variable of interest EOx
is affected by some periodicity (figure 6) in order to avoid such a specific trend. A spectral
analysis is thereby performed, using SPSS (Tukey-Hamming method Brillinger, 2002). Such a
spectral analysis may be done when the number of data points is a power of 2. Therefore, in
order to have the maximum set of points, $2^{15}$ (= 32,768 cases), the first 2 661 data points were
deleted, whence shortening our dataset to the 14/4/1912 - 31/12/2001 interval.

The major peak (we use a $log_2$ scale in order to slightly enhance the x-axis) is about half a year
(183 days or a little bit more than 26 weeks). The next largest peaks occur ca. 372 days (ca. 53
weeks), four months (120 days) and one month (30 days); such "periods" are quite similar to
those found in (Cancho-Candela et al., 2007).

   \subsection{Econometric modelling}
   
   As presented in the above paragraphs, the number of conceptions in a period of time was
assumed to be determined by a set of factors, more or less known and more or less
understood. In the current analysis, classical multiple regression models are considered, with
various components and variables measured on a daily time scale. Several versions of the
models were tested starting from the Ògeneral linearized multifactorialÓ model (Andrei and
Bourbonnais, 2008). All factors mentioned in the data and methods section above were initially
included in the models. Based on the goodness of fit of these, taking into account the statistical
significance of the regression coefficients (with a Student t-test), an optimum, model was
obtained and it will only be the only model outlined below. As in other research (e.g., Lam and
Miron, 1996), the best results are obtained when considering the logarithmic variable as a
dependent variable.
   
  \underline{(A) Model for the orthodox population; components and variables:}
  
  In such a case, the dependent variable, $d^{EOx}$ and covariates are: Days before Christmas
(DBC), Days before Easter (DBE), rurality EOx (rEOx), day length (DL), World wars,
through $1stWW$ and $2ndWW$  variables, weekends (WE), trends (long term tendencies, a stable
component, core of the time series) and a stochastic component ($\epsilon$). The variables are
described in a subsection of the Material and methods section above. A formalized version of
the model can be easily written;   after applying a logarithmic transformation, the model reads in obvious notations, $ln(d^{EOx})$ is equal to 
\begin{eqnarray}
= \alpha_0+  \alpha_1\;DBC + \alpha_2\;DBE+ \alpha_3\;ln(DL)     \nonumber\\  +\alpha_4\;rEOx+ \alpha_5\; 1stWW +
\alpha_6\; 2ndWW + \alpha_7\; WE + \alpha_8\; (trend)^{2} +\epsilon
  \end{eqnarray} \label{eq11}
  
     \underline{(B) Model for the non-orthodox population; components and variables:}
     
      with almost the
same explanatory variables but instead having the dependent variable, $d^{NOx}$, i.e.  the
normalized number of non-orthodox conceptions. Moreover, instead of covariates Òshare of
orthodox in the rural areaÓ, the Òshare of non-orthodox in the rural areaÓ is introduced (rNOx), which may be more suggestive. Thus,   in obvious notations, $ln(d^{NOx})$ is equal to 
\begin{eqnarray}
=  \beta_0+  \beta_1\;DBC + \beta_2\;DBE+ \beta_3\;ln(DL)   \nonumber\\ +\beta_4\;rNOx+ \beta_5\; 1stWW +
\beta_6\; 2ndWW + \beta_7\; WE + \beta_8\; (trend)^{2} +\epsilon
  \end{eqnarray} \label{eq12}
  
          \begin{figure}
 \centering 
 \includegraphics [height=8.0cm,width=10.0cm]{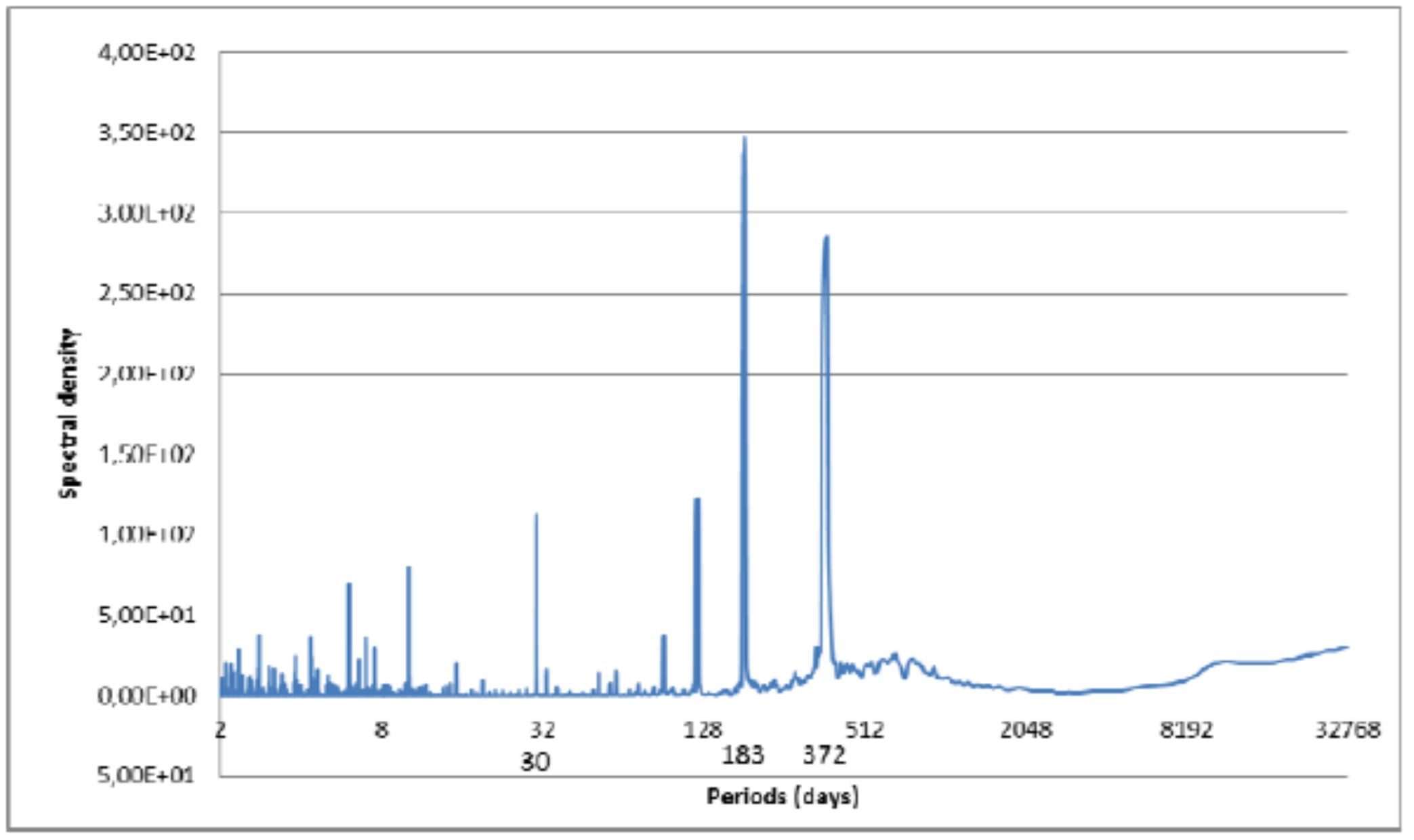}
%Plot10UEFAlili2f.pdf
%  \includegraphics [height=20.0cm,width=16.0cm]{Fig1Plot2colallteams2fits.pdf} \end{figure}
 \caption   {  
Fast Fourier Transformation (FFT) for the $d^{EOx}$.
 This dependent variable in the regression models is calculated by dividing
the daily fraction of births by the daily expected (uniform) distribution across the year. }    \label{figure6}
 \end{figure}  
 
           \begin{figure}
 \centering 
 \includegraphics [height=8.0cm,width=10.0cm]{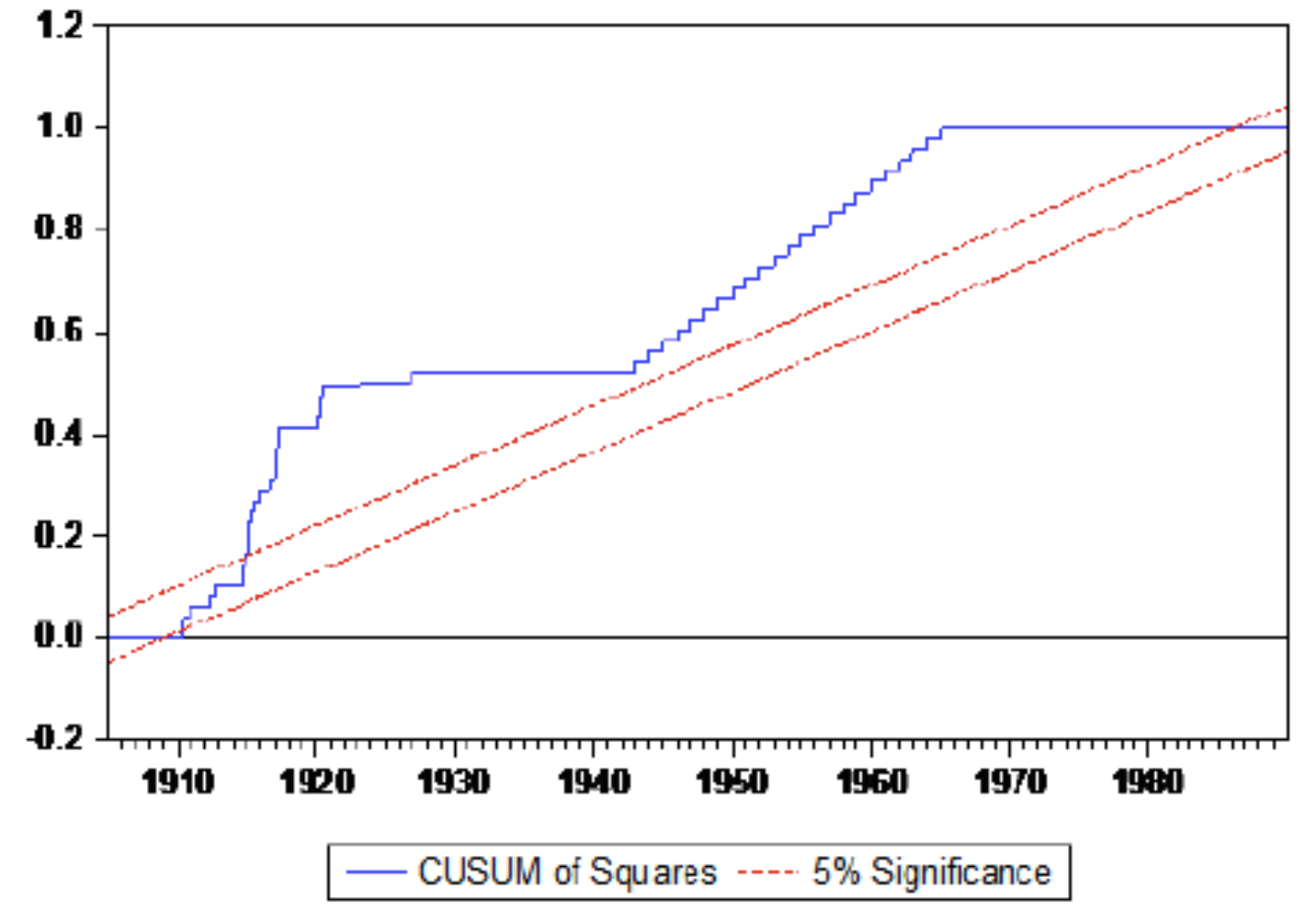}
%Plot10UEFAlili2f.pdf
%  \includegraphics [height=20.0cm,width=16.0cm]{Fig1Plot2colallteams2fits.pdf} \end{figure}
 \caption   { Breaking points identification using CUSUM SQ test }    \label{figure7}
 \end{figure}

    \underline{(C) Structural break identification:}
    
    Since the series covers a very long time interval (97 years=35,429 days), parameters might vary
somewhat. Perhaps as a result the overall model (M0, presented in Table 1) appears not to be
very efficient in that some variables such as the Christmas fast seem to be insignificant. As in
(Cancho-Candela et al., 2007), a segmentation of the long time span into shorter sub-periods
seems an interesting procedural test.
    
In order to do so, the CUSUM SQ (Young, 2011) test was applied to the overall estimated
model (M0). The CUSUM SQ graphic, presented in Figure 7 suggests that a 5\% level of
significance multiple changes is likely during the analyzed timespan. It can be observed that the breaking points are ca. 1911, 1918; 1947, 1966, 1990. Consequently,
there are 6 suggested sub-periods of stability for the models (Mi): 1. 1905-1910 (M1); 2. 1911-
1918 (M2); 3. 1919-1946 (M3); 4. 1947-1966 (M4); 5. 1967-1989 (M5); 6. 1990-2001 (M6). Each break point is in fact well related to a social/ political/historical event: e.g., during the period for
1907-1910, there were peasants' uprisings; at the end of the second and third sub-period, there
were events which determined the decrease of the number of births: the First World War and
the Second World War respectively. Also, this observation suggests how to specify the trend
component of the models. Therefore, the econometric models were reworked for each of the
sub-periods taking into account these remarks: (i) for model M1 a non-linear (let us take it
squared, within usual series expansion approximations) trend of births during the entire period
is introduced; (ii) for model M2 and M3, similar squared trends can be also assumed.; (iii) the
same assumptions are introduced for models M4 and M5. According to national reports, %such as ($http://www.insse.ro/cms/files/publicatii/Evolutia\%20natalitatii\%20si\%20fertilitatii\%20in\%20Romania_{-}n.pdf$, retrieved in November 2014)    
during the period of 1956-1966, the nativity rate
decreased dramatically due to many factors such as: abortion liberalization, better access to
work and education for women; moreover, between 1985 and 1990, the social condition of
Romanians significantly decreased. The Decree of Ceausescu outlawing abortion (1st October
1966) had effects only for a short period of time; (iv) finally, for the 6th model, the trend in the
number of babies conceived and born decreased according to many factors, like abortion
liberalization (992 000 such operations occur in 1990), but also post-communist socio-economic
crisis, migration etc.
    
    After applying estimations using SPSS methodology, the optimized results are given in Table 2
and Table 3.
    
    After performing a F test - ANOVA it has been found that the model for every sub-period for
EOx and NOx population are valid and statistically significant. The proportion of
variance (R2) explained for the dependent variable is between 12.7\% (M0 and M5) and 32.3\%
(M3), - values similar to other studies (Friger et al., 2009). Almost all regression parameters are
statistically significant (from a Student t-test) with slightly different degrees of significance
(most of them having the p value lower than 0.01). As expected, the model performed on
NOx population, registered levels of R2 lower than the previous ones. The maximum
deviation occurs for M6 (10.5\%) and the minimum for M2 (2.3\%). Regression coefficients in this
case fail to be slightly more often significant.
    
    In addition, the sensitivity of the statistical models was tested by taking into consideration a
gestation duration shorter by one and two weeks, respectively. The models outcomes in these
scenarios are found to be quite similar to the presented ones.
    
   \section{Discussion and Conclusions} \label{sec:conclusions}

Romanians following Eastern Orthodox religious beliefs or rules, i.e. the majority of religious
persons in Romania, are taught to avoid sexual intercourse and to abstain from particular food
and drinks during fasting periods. The results of our time series analysis on birth rates show
different behaviors during the major fasts. Conceptions, during Lent Fast, are consistent with
religious constraints. The negative value of the coefficient (DBE) for Days to Easter during the
Fast shows that as the Holiday gets closer, the estimated number of conceptions significantly
decreases. It is worth mentioning here that the Orthodox religion grants some exception
periods to believers for reasons mainly linked to health. In these Ògrace periodsÓ, there might
be increased conception. It is important to note that the standard absolute value of the
coefficient of the variable which measures the impact of Lent fast on conception (DBE) remains
relatively stable in [-0.25; -0.35] for all sub-periods. This points to the continuity and
persistence of the Eastern Orthodox faith in general, and particularly regarding interdictions of
Lent fasting throughout the almost 100 years of this study. This continuous adhesion to faith is
also highlighted by the values estimated in the second model (Table II) for the Non-Orthodox
sub sample (NOx) where the estimated coefficient of the variable has reasonable values,
though half [-0.04;-0.19] that for the EOx population.
On the other hand, the Nativity Fast no-sex constraint does not seem to have a great
significance for Orthodox Romanians nowadays: the coefficient of the variable (DBC) which
measures this impact is positive after the 1st World War, showing that as the Christmas day gets
closer, the number of conceptions increases significantly. If in the case of the Lent Fast the
coefficient remains stable in time, in the case of the Nativity Fast, the standard value is
increasing in time, showing that when approaching the Nativity date, the population practices a
more liberal lifestyle, obeying fewer constraints during the fasting period. For example, in the
period 1990-2001 the value of the coefficient is approximately three times higher than the one
estimated for the period before the Second World War. This can be likely explained from a rural
habit: weddings are traditionally performed in rural Romania in the last part of autumn, after
the harvest (Trebici, 1979). Moreover, the fermentation process of the new wine is stopped,
thus rendering the liquid more drinkable, in particular at weddings. It has been shown (Trebici,
1979) that usually about 9-10 months later, the first baby of a new couple is delivered. Another
point is based on the psychological idea that the Christmas holiday is a more joyful event
compared to Easter (in agreement with others, like Seive (1985)). We can also propose as a
cause that, in rural Romania, there are fewer household duties (see Lam and Miron, 1991;
tangential Becker, 1991) occurring outside the house during the Nativity fast than during spring
Lent. And lastly, the day length (and of course night length) needs to be mentioned as there
was a lack of electricity until the middle of the 20th century and also a lack of other indoor
sources of entertainment, thus increasing the likelihood of sexual encounters at those times.

Therefore, the proposed models highlight the impact of religiously controlled fasting periods on
conception, besides the existence of other factors with persistent or temporary influence. For
instance, the day length variable reveals the significant influence of several factors such as: the
regularity of daily activities determined by the season and the temperature (these two factors
were included in the initial models and eliminated due to the multicolinearity effect). With one
exception (M1), our analysis reveals a photoperiodicity related cause for baby conception as
found for European Countries like France, England, Sweden (Lam and Miron, 1994) on monthly
data. Of course, due to multicolinearity, the photoperiodicity could have only an apparent
effect as a proxy for factors like the temperature, outside household duties or the lack of
available inside sources of entertainment.

Weekend is a factor inserted into our models that tends to function as proxy for holidays, and
more permissive sexual activity. Previous research (Bobak and Gjonca, 2001) suggests that
during holidays, the frequency of conception seems to be higher indeed. Our data analysis
generally agrees with these findings. Therefore, for EOx population, weekends seem to have
a positive impact on conception with the exception of M4. Furthermore, the same influence is
reflected on NOx population also until 1966 (models M1 to M4). Thereafter, weekends are
shown to have a negative influence.

It also needs to be stated that despite other researchers' findings (Jongbloet, 1983), the
proportion of illuminated moon disc was found to be insignificant from a statistical view point
(along a Student t-test).

Furthermore, the values of the regression coefficients specific to the Orthodox persons in the
rural environment, illustrate that behavioral patterns in these geographical areas are much
more prone to seasonal elements in child conception as compared to urban areas, where these
differentiations are diminished.

The trend of the series for each analyzed sub-period confirms the expectations stated above.
Thus, in each major timespan here analyzed, there are years of maximum increases followed by
a descending evolution. The only exception to this rule are the years between 1948-1966, when
the maximum points of the number of conceived babies are at the ends of the interval. On the
one hand, this could be determined by the ÒnecessityÓ to cover the disaster caused by the war;
on the other hand, it shows the significant impact of Ceausescu's decree.

Finally, the splitting of the populations into EOx and NOx categories is proven relevant.
The model for all sub-periods analyzed is much weaker (from econometrics viewpoint) for the
NOx population which does not have Nativity or Lent fasts, or if it does, with time delays
versus the EOx population. Moreover, the standardized coefficient that measures the impact of Lent Fast on conception has in general the largest absolute value. It is followed in
absolute value by the rurality (rEOx) coefficient, both values strengthening the value of the
orthodox tradition and faith, factors which are underestimated or missing in previous human
conception researches.

In brief, the primary result of the current investigation is that the non-uniformity of the
conception of babies, in Romania, for the last century, across the year is affected not merely by
seasonality, but by a double filter: season and religion. %The econometric-based model which we propose is in full agreement with the features, and can serve for discussing  other   cases available to sociophysics concerns.  

    \vskip0.5cm
 {\bf Acknowledgements} 
 
  During the long gestation period for this paper several relatives, friends, colleagues of the
authors provided valuable feed-back, suggestions, comments: Robert Ancuceanu, Tudorel
Andrei, Radu Chirvasuta, Gurjeet Dhesi, Victor Dragota, Anca Herteliu, Roxana Herteliu-Iftode,
Alexandru Isaic-Maniu, Ionel Jianu, Michael S. Jones, Adrian Pana, Gheorghe Peltecu. And last
but not least Dr. David Berman (Iowa University) reviewed the paper for its grammar,
vocabulary and style content. He has made also many very interesting suggestions improving
the scientific content.
 
 This paper is part of scientific activities in COST Action  TD1210 'Analyzing the
dynamics of information and knowledge landscapes'. This work by CH was co-financed by the
European Social Fund through the Sectorial Operational Programme Human Resources
Development 2007-2013, project number POSDRU/1.5/S/59184 Performance and excellence in
postdoctoral research in Romanian economics science domain

\end{document}